\documentclass{article}
\usepackage{stmaryrd}
\usepackage{amssymb}
\usepackage{graphics}
\usepackage{epsfig}
\usepackage{a4wide}
\usepackage{cite}
\usepackage{color}
\usepackage{graphicx}
\usepackage{dcolumn}
\usepackage{bm}
\usepackage{indentfirst}

\begin{document}

\title{Minor ion heating in spectra of linearly and circularly polarized Alfv\'{e}n waves: Thermal and non-thermal motions associated with perpendicular heating}
\author{Chuanfei Dong$^\mathrm{a,b}$\thanks{%
dcfy@umich.edu} \\
{$^\mathrm{a}${\small \emph{Department of Atmospheric, Oceanic and Space
Sciences, University of Michigan,}}}\\
{\small \emph{\ Ann Arbor, MI 48109, U.S.A.}}\\
{$^\mathrm{b}${\small \emph{Los Alamos National Laboratory, Los Alamos, NM
87544, U.S.A. }}}}
\date{}
\maketitle

\begin{abstract}
Minor ion (such as $He^{2+}$) heating via nonresonant interaction with spectra of linearly and circularly polarized Alfv\'{e}n waves (LPAWs and CPAWs hereafter) is studied. The obtained analytic solutions are in good agreement with the simulation results, indicating that newborn ions are heated by low-frequency Alfv\'{e}n waves with finite amplitude in low-beta plasmas such as the solar corona. The analytic solutions also reproduce the preferential heating of heavy ions in the solar wind. In the presence of parallel propagating Alfv\'{e}n waves, turbulence-induced particle motion is clearly observed in the wave (magnetic field) polarized directions. After the waves diminish, the newborn ions are heated, which is caused by the phase difference (randomization) between ions due to their different parallel thermal motions. The heating is dominant in the direction perpendicular to the ambient magnetic field. The perpendicular heating, $\eta=(T_{i\perp}^R-T_{i0\perp}^R)/T_{i0\perp}^R$ (where $T_{i0\perp}^R$ and $T_{i\perp}^R$ are the perpendicular temperature of species $i$ before and after genuine heating, respectively), in the spectrum of CPAWs is a factor of two stronger than that of LPAWs. Moreover, we also study the effect of field-aligned differential flow speed of species $i$ relative to $H^+$, $\delta v_{ip} = (\mathbf{v_i}-\mathbf{v_p}) \cdot \mathbf{B} / |\mathbf{B}|$ (where $\mathbf{v_i}$ and $\mathbf{v_p}$ denote vector velocities of the $H^+$  and species $i$, respectively), on the perpendicular heating. It reveals that large drift speed, $v_d=\delta v_{ip}$, has an effect on reducing the efficiency of perpendicular heating, which is consistent with observations. 
  
\end{abstract}

\vskip 15mm

\renewcommand{\theequation}{\arabic{section}.\arabic{equation}} %
\renewcommand{\thesection}{\Roman{section}.}

\vfill \eject

\baselineskip=0.32in

\section{INTRODUCTION}

In collisionless space environment, wave-particle interactions can dominate over fluid or collisional processes, resulting in, e.g., temperature anisotropy, which is a common phenomenon that has been observed in the solar corona and solar wind for a long time \cite{Marsch,Marsch1,An2}. Meanwhile, \emph{Wind} spacecraft \cite{JK1,JK2} and Helios \cite{Marschhe,SB1,SB2} mission has been offering great observational results showing that heavy ion such as $\alpha$ particles are preferentially heated perpendicular to the background magnetic field direction in the solar wind. Alfv\'{e}n/ion-cyclotron waves have long been considered to play a crucial role in heating of plasma in the solar wind and the solar corona\cite{Marsch}. Studies of ion heating by Alfv\'{e}n/ion-cyclotron waves through resonant heating mechanism, using different theoretical models, date back more than a decade\cite{res2,res3,Marsch,Marschhe}, where the Alfv\'{e}n wave frequency is always comparable to the ion cyclotron frequency. Later, Isenberg \emph{et al.}\cite{PI} proposed a second-order Fermi mechanism to explain the preferential perpendicular heating based on multiple cyclotron resonances with counterpropagating field-aligned ion-cyclotron waves. This model has become very popular recently since the comparison of the model predictions to observations from the \emph{Wind} spacecraft reveals excellent agreement\cite{JK2}. In Ref.\cite{JK2}, Kasper \emph{et al.} started with a cold plasma dispersion relation and pointed out their analysis can be improved by adopting a more realistic dispersion relation that accounts for finite temperature and multiple species\cite{JK2,PSM}.

On the other hand, it is not clear if high-frequency Alfv\'{e}n/ion-cyclotron waves can be produced in sufficient quantities to explain the levels of proton heating that are observed. According to observations, it is unlikely that the cascade of both non-compressive MHD fluctuations and compressive magnetosonic waves are a significant source of high-frequency ion-cyclotron fluctuations in the near-Earth solar wind\cite{Ben}. Dissipation of low-frequency Alfv\'{e}n turbulence, rather than high frequency ion-cyclotron waves, offers an alternative explanation for perpendicular ion heating in the solar wind, indicating heating occurs at $\omega \ll {\Omega}_{i}$ ($\omega$ is the frequency of the Alfv\'{e}n waves and ${\Omega}_{i}$ denotes the gyrofrequency for ion species labeled \emph{i}). Many viable theories requiring the dissipation of waves through low frequency Alfv\'{e}n wave-particle interaction have been proposed, such as stochastic heating\cite{chenPOP,Ben,Ben1,Xia,DV2,WangGRL,LuApJ}, nonresonant wave-particle scattering\cite{dongPOP,dongPOP2,luCPB,luPOP,wangPRL,SAPL}, parametric instabilities\cite{jaaprl2008,jaaprl2009} and superadiabatic acceleration\cite{YAPL}. Results from simulations showed a reasonable agreement with those obtained from analytical theories. Among various proposed models, the stochastic heating mechanism\cite{chenPOP,Ben,Ben1,Xia,DV2,WangGRL,LuApJ} has become more and more popular. After extending previous theories of stochastic ion heating to account for the motion of ions along the magnetic field $\mathbf{B}$, Chandran \emph{et al.}\cite{Ben} shown that it can reproduce most of the observational feature of the ion temperature by \emph{Wind} spacecraft. However, the stochastic heating model occurs only when certain threshold condition is satisfied, which usually requires large amplitude Alfv\'{e}n waves\cite{Ben1,WangGRL,chenPOP,LuApJ}. Besides, it takes several thousands to tens of thousands ion gyroperiods to reach the asymptotic results\cite{Ben1,LuApJ,WangGRL}, which is relatively long to the time scale needed for nonresonant wave-particle scattering\cite{wangPRL,dongPOP}. Although the heating efficiency of nonresonant wave-particle scattering is low with respect to the stochastic heating\cite{WangGRL,LuApJ}, it does not have the limitation of threshold condition, and usually finishes within one cyclotron period \cite{wangPRL,dongPOP}. Therefore, nonresonant wave-particle scattering may also be important in some circumstances.

Meanwhile, the shape of velocity distribution functions (VDFs) due to wave activity has been widely discussed recently\cite{wangPRL,wuPRL,wangPOP,wuPOP1,DV,dongPOP2013,Fisk,YN}. The deformation of VDFs with respect to a Maxwellian shape caused by the presence of wave field forces (or their spectra) may mimic genuine heating. The wave field forces would lead to a velocity spread after appropriately averaging over the wave effects, which results in an effective non-thermal broadening of the VDFs similar to real heating process \cite{DV}. Therefore, this process has been named as ``pseudoheating'' and the corresponding temperature as ``apparent temperature'' \cite{wangPOP,DV,dongPOP2013}. The characteristics of wave effect on the VFDs are reversible and non-dissipative. Thus, it does not represent the genuine heating and the correlated particle motion is nonthermal. Interestingly, pseudoheating can mimic the perpendicular heating as described above, especially when the wave amplitude is large\cite{DV,dongPOP2013}. In the case of finite amplitude low-frequency Alfv\'{v}e wave, both nonresonant wave-particle scattering and pseudoheating could happen within one ion gyroperiod, thus distinguishing pseudoheating from real heating is important \cite{DV,dongPOP2013,wangPOP,wuPRL,wuPOP1,YN}. All effects mentioned above, and many others, lead to a superabundance of phenomena involving particles and fields.

In this paper, we show that the low-frequency Alfv\'{e}n waves propagating parallel to the background magnetic field can heat ions in low-$\beta$ plasmas (the plasma $\beta $ is the ratio of thermal pressure to magnetic pressure). Both LPAWs and CPAWs are considered. We find that the heating processes under these two polarized conditions are different. The comparison of the simulation results clearly shows the difference between real and pseudo heating. The organization of this paper is as follows: We derive and discuss the analytic results of non-resonant ion heating by spectra of both CPAWs and LPAWs in Sec. II. In Sec. III, test particle simulation results are presented and discussed based on the comparison of two polarized cases. The simulation results are in good agreement with the analytic solutions. Furthermore, we present the effect of drift (or differential flow) speed on the $\alpha$ particle perpendicular heating. In the last section, conclusions are summarized.

\vskip 10mm

\section{ANALYTIC THEORY OF NON-RESONANT WAVE-PARTICLE SCATTERING}

We consider that Alfv\'{e}n waves have a spectrum and propagate along the background magnetic field $\mathbf{B_0}=B_0 \mathbf{i}_{z}$. The dispersion relation can be described as $\omega =k{v}_{A}$ (${v}_{A}$ is the Alfv\'{e}n speed, $\omega$ and $k$ are the wave angular frequency and the wave number, respectively). Two different polarized cases are studied. Without loss of generality, we consider left-hand CPAWs, and LPAWs with $\delta B_y^{lnr}$. The wave magnetic field, $\delta {\mathbf{B}}_{w}$, can be expressed as 
\begin{equation}
\delta {\mathbf{B}}_{w}^{lnr}=\sum_{k}{B}_{k}\cos {\phi }_{k}{\mathbf{i}}%
_{y},  \label{lnr}
\end{equation}%
\begin{equation}
\delta {\mathbf{B}}_{w}^{cir}=\sum_{k}{B}_{k}\left( \cos {\phi }_{k}{\mathbf{%
i}}_{x}-\sin {\phi }_{k}{\mathbf{i}}_{y}\right) ,  \label{cir}
\end{equation}%
where $\delta {\mathbf{B}}_{w}^{lnr}$ and $\delta {\mathbf{B}}_{w}^{cir}$
represent linearly and circularly polarized magnetic perturbations, respectively. 
The electric field, $\delta {\mathbf{E}}_{w}$, is
\begin{equation}
\delta {\mathbf{E}}_{w}=-v_{A}{\mathbf{i}}_{z}\times \delta {\mathbf{B}}_{w},
\label{e-fileld}
\end{equation}%
where ${\mathbf{i}}_{x}$, ${\mathbf{i}}_{y}$ and ${\mathbf{i}}_{z}$ are unit
directional vectors, $\phi _{k}=k(v_{A}t-z)+\varphi _{k}$ denotes the wave
phase, and $\varphi _{k}$ is the random phase of mode $k$. The equation of
motion (EOM) for an ion is described by%
\begin{equation}
m_{i}\frac{d\mathbf{v}}{dt}=q_{i}\left( \delta {\mathbf{E}}_{w}+\mathbf{v}%
\times ({\mathbf{B}}_{0}+\delta {\mathbf{B}}_{w})\right) ,~~~\frac{d\mathbf{r%
}}{dt}=\mathbf{v},  \label{vtot}
\end{equation}%
where $\mathbf{v}$ is the ion velocity, $m_{i}$ and $q_{i}$ are the ion mass
and charge, respectively.

Let's start with the CPAWs. Defining ${v}_{\perp }={%
v}_{x}+i{v}_{y}$, ${v}_{\parallel }={v}_{z}$ and $\delta B_{\omega
}=\sum_{k}B_{k}e^{-i\phi _{k}}$; we are left with 
\begin{equation}
\frac{dv_{\perp }}{dt}+i\Omega _{0}v_{\perp }=i(v_{\parallel
}-v_{A})\sum_{k}\Omega _{k}e^{-i\phi _{k}},  \label{cvperp}
\end{equation}%
\begin{eqnarray}
\frac{dv_{\parallel }}{dt} &=&-v_{x}\sum_{k}\Omega _{k}\sin \phi
_{k}-v_{y}\sum_{k}\Omega _{k}\cos \phi _{k}  \nonumber \\
&=&-Im(v_{\perp }\sum_{k}\Omega _{k}e^{i\phi _{k}}),~~~\frac{dz}{dt}%
=v_{\parallel },  \label{cvpara}
\end{eqnarray}%
where $\Omega _{0}=q_{i}B_{0}/m_{i}$ (the ion gyrofrequency), $\Omega
_{k}=q_{i}B_{k}/m_{i}$. $Im()$ denotes the imaginary part of its argument.
As a first-order approximation, we assume $v_{\parallel }\approx
v_{\parallel }(0)$, where $v_{\parallel }(0)$ is the particle's initial
parallel velocity. The approximation is valid when $\sum_{k}B_{k}/B_{0}$ is small enough and the
frequencies of the Alfv\'{e}n wave are sufficiently low to ensure that $\left\vert 
\Omega _{0}\right\vert \gg \left\vert k\left( v_{\parallel }(0)-v_{A}\right) \right\vert $. With
the initial condition $v_{\perp }=v_{\perp }(0)$, $v_{\parallel
}=v_{\parallel }(0)$ and $z=z(0)$, the solution of ordinary differential
equation, Eq.(\ref{cvperp}), can be written as: 
\begin{eqnarray}
v_{\perp } &=&v_{\perp }(0)e^{-i\Omega _{0}t}-v_{A}\frac{\sum_{k}\Omega _{k}%
}{\Omega _{0}}e^{-ik(v_{A}t-z)-i\varphi _{k}}+v_{A}\frac{\sum_{k}\Omega _{k}%
}{\Omega _{0}}e^{i\left[ kz(0)-\varphi _{k}\right] }e^{-i\Omega _{0}t} 
\nonumber \\
&=&v_{\perp }(0)e^{-i\Omega _{0}t}-v_{A}\frac{\sum_{k}B_{k}}{B_{0}}%
e^{-ik(v_{A}t-z)-i\varphi _{k}}+v_{A}\frac{\sum_{k}B_{k}}{B_{0}}e^{i\left[
kz(0)-\varphi _{k}\right] }e^{-i\Omega _{0}t},  \label{csolprep}
\end{eqnarray}%
substituting Eq.(\ref{csolprep}) into Eq.(\ref{cvpara}), we obtain 
\begin{equation}
v_{\parallel }=v_{\parallel }(0)+v_{A}\frac{\sum_{k}B_{k}^{2}}{B_{0}^{2}}%
\times \left\{ 1-\cos \left[ \Omega _{0}t-kv_{A}t+kv_{\parallel }(0)t\right]
\right\} ,  \label{csolpara}
\end{equation}%
where we use the approximation that $\Omega _{0}-k[v_{A}-v_{\parallel
}(0)]\approx \Omega _{0}$, $v_{A}\gg \left\vert v_{\parallel }(0)\right\vert  $, $v_{A}\gg
\left\vert  v_{\perp }(0)\right\vert  $ and $z=z(0)+v_{\parallel }(0)t$. The solution is
valid for a long time interval as discussed in Ref.\cite{wangPRL}. The
transverse motion of the particle consists of three terms: the gyromotion of
the particle in the background magnetic field $\mathbf{B_{0}}$; the transverse motion
due to the electric field of the Alfv\'{e}n waves; and the modification of
the gyromotion due to the existence of the Alfv\'{e}n waves. In order to
distinguish the genuine heating from pseudoheating later, we calculated the
overall average transverse and parallel velocities at position $z$ in the
particles' mean-velocity frame as follows: 
\begin{eqnarray}
U_{\perp } &=&-v_{A}\frac{\sum_{k}B_{k}}{B_{0}}e^{-ik(v_{A}t-z)-i\varphi
_{k}}  \nonumber \\
&&+\frac{1}{\sqrt{\pi }v_{th}}v_{A}\int_{-\infty }^{\infty }\frac{%
\sum_{k}B_{k}}{B_{0}}e^{ik\left[ z-v_{\parallel }(0)t\right] -i\varphi
_{k}}e^{-i\Omega _{0}t}e^{-\left[ \frac{v_{\parallel }(0)}{v_{th}}\right]
^{2}}dv_{\parallel }(0)  \nonumber \\
&=&-v_{A}\frac{\sum_{k}B_{k}}{B_{0}}e^{-ik(v_{A}t-z)-i\varphi
_{k}}+A_{k}v_{A}\frac{\sum_{k}B_{k}}{B_{0}}e^{ikz-i\varphi _{k}}e^{-i\Omega
_{0}t},  \label{cbperp}
\end{eqnarray}%
where $A_{k}=\left( 1/\sqrt{\pi }\right) \int_{-\infty }^{\infty }\cos
\left( kv_{th}tx\right) e^{-x^{2}}dx=e^{-k^{2}v_{th}^{2}t^{2}/4}$ and the particle initial thermal speed $v_{th}=\left( 2k_{B}T_{0}/m_{i}\right) ^{1/2}$ (by notation $x=v_{\parallel }(0)/v_{th}$ and $T_{0}$ is the ion initial temperature). Here we assume that all particles initially satisfy a Maxwellian velocity distribution. This average transverse velocity illustrates the ion pickup by the Alfv\'{e}n waves. It is obvious that $v_{\parallel }$ can be considered as $v_{\parallel }(0)$ if the amplitude of the Alfv\'{e}n wave is sufficienctly small. By adopting the same integration, the average parallel velocity at $z$ can be written as 
\begin{eqnarray}
U_{\parallel } &=&\frac{1}{\sqrt{\pi }v_{th}}v_{A}\int_{-\infty }^{\infty }%
\frac{\sum_{k}B_{k}^{2}}{B_{0}^{2}}\times \left\{ 1-\cos \left[ \Omega
_{0}t-kv_{A}t+kv_{\parallel }(0)t\right] \right\} e^{-\left[ \frac{%
v_{\parallel }(0)}{v_{th}}\right] ^{2}}dv_{\parallel }(0)  \nonumber \\
&=&v_{A}\frac{\sum_{k}B_{k}^{2}}{B_{0}^{2}}\left[ 1-A_{k}\cos \left( \Omega
_{0}t-kv_{A}t\right) \right] ,  \label{cbupara}
\end{eqnarray}

To simplify the notation, we define $\delta B_{w}^{2}/
B_{0}^{2}=\sum_{k}B_{k}^{2}/B_{0}^{2}$ and $\delta B_{w}^{4}/B_{0}^{4}=\left(\sum_{k}B_{k}^{2}/B_{0}^{2}\right)^2$ hereafter. Since we know both the perpendicular velocity $v_{\perp}$ and averaged transverse velocity $U_{\perp} $, the perpendicular temperature is 
\begin{eqnarray}
T_{cir\perp }^{R} &=&\frac{m_{i}}{2k_{B}\sqrt{\pi }v_{th}}\int_{-\infty
}^{\infty }\left\vert v_{\perp }-U_{\perp }\right\vert ^{2}e^{-\left[ \frac{%
v_{\parallel }(0)}{v_{th}}\right] ^{2}}dv_{\parallel }(0)  \nonumber \\
&\simeq &T_{0}+\frac{m_{i}}{2k_{B}}\left[ \frac{v_{A}^{2} \sum_{k}B_{k}^{2}}{B_{0}^{2}} 
\left( 1-A_{k}^{2}\right) \right]   \nonumber \\
&= &T_{0}\left[ 1+\frac{m_{i}}{m_{p}}\frac{\delta B_{w}^{2}}{\beta
_{p}B_{0}^{2}}\left( 1-A_{k}^{2}\right) \right] ,  \label{cTRperp}
\end{eqnarray}
where $\beta _{p}=2\mu _{0}n_{p}k_{B}T_{0}/B_{0}^{2}$ and $m_{p}$ are the proton plasma beta and proton mass, respectively. The superscript $R$ indicates the temperature is achieved by real heating. The increase of the perpendicular temperature is due to the initial random velocities of particles in the parallel direction. According to the third term on the Eq.(\ref{csolprep}), particles at position $z$ will have different velocities after time $t$, causing by the phase difference between particles. As a result, a velocity dispersion is produced and ions are heated in the perpendicular direction. The parallel temperature can be expressed as 
\begin{eqnarray}
T_{cir\parallel }^{R} &=&\frac{m_{i}}{k_{B}\sqrt{\pi }v_{th}}\int_{-\infty
}^{\infty }\left\vert v_{\parallel }-U_{\parallel }\right\vert ^{2}e^{-\left[
\frac{v_{\parallel }(0)}{v_{th}}\right] ^{2}}dv_{\parallel }(0)  \nonumber \\
&\simeq &T_{0}\left\{ 1+\frac{2m_{i}}{m_{p}}\frac{\delta B_{w}^{4}}{\beta
_{p}B_{0}^{4}}\left[ \left( B_{k}-A_{k}^{2}\right) \cos ^{2}\left( \Omega
_{0}t-kv_{A}t\right) +C_{k}\sin ^{2}\left( \Omega _{0}t-kv_{A}t\right) %
\right] \right\} ,  \label{cTRpara}
\end{eqnarray}
where $B_{k}=\left( 1/\sqrt{\pi }\right) \int_{-\infty }^{\infty }\cos
^{2}\left( kv_{th}tx\right) e^{-x^{2}}dx=0.5+0.5e^{-k^{2}v_{th}^{2}t^{2}}$
and $C_{k}=\left( 1/\sqrt{\pi }\right) \int_{-\infty }^{\infty }\sin
^{2}\left( kv_{th}tx\right) e^{-x^{2}}dx=0.5-0.5e^{-k^{2}v_{th}^{2}t^{2}}$.
When $t\rightarrow \infty $, $A_{k}\rightarrow 0$, $B_{k}\rightarrow 0.5$, and $
C_{k}\rightarrow 0.5$.Thus the asympotic average transverse and parallel
velocities, average perpendicular and parallel temperatures are 
\begin{eqnarray}
U_{\perp } &=&-v_{A}\frac{\sum_{k}B_{k}}{B_{0}}e^{-ik(v_{A}t-z)-i\varphi
_{k}},  \label{cbuperp1} \\
U_{\parallel } &=&v_{A}\frac{\sum_{k}B_{k}^{2}}{B_{0}^{2}},  \label{cbupara1}
\\
T_{cir\perp }^{R} &=&T_{0}\left( 1+\frac{m_{i}}{m_{p}}\frac{\delta B_{w}^{2}%
}{\beta _{p}B_{0}^{2}}\right) ,  \label{cTRperp1} \\
T_{cir\parallel }^{R} &=&T_{0}\left( 1+\frac{m_{i}}{m_{p}}\frac{\delta
B_{w}^{4}}{\beta _{p}B_{0}^{4}}\right) ,  \label{cTRpara1}
\end{eqnarray}
Compared Eq.(\ref{cTRperp1}) with Eq.(\ref{cTRpara1}), it is  clear that the ion heating is dominant in the perpendicular direction.
Due to the existence of shear Alfv\'{e}n wave fluctuations in the perpendicular direction, we will incorporate the wave effect to the perpendicular heating in the following derivation. Given the assumption that the characteristic spatial scale of the system is much larger than the typical Alfv\'{e}n wavelength, we can take an ensemble average of $|v_{\perp}|^2$ [see Eq.(\ref{csolprep})]. Similar to the procedure used in Refs.\cite{dongPOP,wangPRL}, we obtain the expressions of the asympotic kinetic temperature for time scale larger than $\pi /(kv_{th})$ as follows, 
\begin{eqnarray}
T_{cir\perp }^{A+R} &=&\frac{1}{2k_B}m_i \langle |v_{\perp}|^2 \rangle  \nonumber \\
&\simeq &T_{0}+\frac{m_iv_{A}^{2}}{k_B}\frac{\delta B_{w}^{2}}{%
B_{0}^{2}}  \nonumber \\
&\simeq &T_{0}\left( 1+\frac{2}{{\beta }_{p}}\frac{m_{i}}{m_{p}}\frac{\delta
B_{w}^{2}}{B_{0}^{2}}\right) \label{cTKperp},
\end{eqnarray}
where the superscript $A$ represents the apparent temperature\cite{wangPOP,DV,dongPOP2013,YN} that results from the pseudoheating. $T_{cir\perp }^{A+R}$ is the temperature achieved by both real and pseudo heating. The comparison between Eq.(\ref{cTRperp1}) and Eq.(\ref{cTKperp}) reveals that the temperature $T_{cir\perp }^{A+R}$ is higher than $T_{cir\perp }^{R}$ due to the presence of wave field forces (or their spectra). By employing the formula in Ref.\cite{Landi,NT2}, we define the total thermal speed $v_{th}^{tot}$ as the combination of thermal speed $v_{th}^{R}$ associated with genuine heating and non-thermal speed $\xi $ correlated to the pseudoheating.
\begin{eqnarray}
v_{th}^{tot} &=&\sqrt{v_{th}^{R~2}+\xi ^{2}}  \nonumber \\
&=&\sqrt{\frac{2k_{B}T_{cir\perp }^{A+R}}{m_{i}}},
\end{eqnarray}%
we get $\xi =v_{A}\delta B_{w}/B_{0}$, which is consistent with Ref. 
\cite{NT2}, where they argued that the non-thermal speed depends on the wave amplitude.

In the case of LPAWs, the corresponding EOM is described by 
\begin{equation}
\frac{dv_{\perp }}{dt}+i\Omega _{0}v_{\perp }=(v_{A}-v_{\parallel
})\sum_{k}\Omega _{k}\cos \phi _{k},  \label{lvperp}
\end{equation}%
\begin{equation}
\frac{dv_{\parallel }}{dt}=v_{x}\sum_{k}\Omega _{k}\cos \phi _{k},~\frac{dz}{%
dt}=v_{\parallel },  \label{lvpara}
\end{equation}

With similar assumptions ($v_{\parallel} \approx v_{\parallel}(0)$, $\left\vert\Omega_0\right\vert \gg \left\vert k\left[v_{\parallel}(0)-v_A\right] \right\vert, \left\vert v_{\parallel}(0)\right\vert \ll v_A, \left\vert v_{\perp}(0)\right\vert \ll v_A$ ) and procedures, by ignoring the terms proportional to $kv_A / \Omega_0$, the solution of ordinary differential equation, Eq.(\ref{lvperp}), is:
\begin{eqnarray}
v_{\perp } &=&v_{\perp }(0)e^{-i\Omega _{0}t}+e^{-i\Omega _{0}t}\int_{0}^{t}\left[
v_{A}-v_{\parallel }\left( 0\right) \right] \sum_{k}\Omega _{k}\cos \left[
k\left( v_{A}-v_{\parallel }\left( 0\right) \right) t-kz\left( 0\right)
+\varphi _{k}\right] e^{i\Omega _{0}t}dt  \nonumber \\
&\simeq &v_{\perp }(0)e^{-i\Omega _{0}t}+iv_{A}\frac{\sum_{k}B_{k}}{B_{0}}\cos %
\left[ \varphi _{k}-kz(0)\right] e^{-i\Omega _{0}t} 
-iv_{A}\frac{\sum_{k}B_{k}}{B_{0}}\cos \left[ \varphi _{k}+kv_{A}t-kz%
\right] ,  \label{lsolperp}
\end{eqnarray}
where $v_{\perp }$ can be decomposed into $v_{x}$ and $v_{y}$ as
follows: 
\begin{eqnarray}
v_{x} &=&v_{x}(0) +v_{A}\frac{%
\sum_{k}B_{k}}{B_{0}}\cos \left[ \varphi _{k}-kz(0)\right] \sin \left(
\Omega _{0}t\right) ,  \label{lsolvx} \\
v_{y} &=&v_{y}(0) -v_{A}\frac{%
\sum_{k}B_{k}}{B_{0}}\cos \left[ \varphi _{k}+kv_{A}t-kz\right]   \nonumber
\\
&&+v_{A}\frac{\sum_{k}B_{k}}{B_{0}}\cos \left[ \varphi _{k}-kz(0)\right]
\cos \left( \Omega _{0}t\right) ,
\end{eqnarray}

The velocity components $v_{x}$ and $v_{y}$, however, are no longer symmetric due to the presence of magnetic perturbation, $\delta {\mathbf{B}}_{w}^{lnr}$, only in $y$ direction. Substituting Eq.(\ref{lsolvx}) to Eq.(\ref{lvpara}), we obtain 
\begin{eqnarray}
\frac{dv_{\parallel }}{dt} &=&\left( v_{x }(0) +v_{A}\frac{\sum_{k}B_{k}}{B_{0}}\cos \left[ \varphi
_{k}-kz\left( 0\right) \right] \sin \left( \Omega _{0}t\right) \right)
\times  \nonumber \\
&&\sum_{k}\Omega _{k}\cos \left[ k\left( v_{A}-v_{\parallel }\left( 0\right)
\right) t-kz\left( 0\right) +\varphi _{k}\right] ,
\end{eqnarray}
given $v_{\perp}(0) \ll v_A $, the solution of the equation, $v_{\parallel }$,  is
\begin{eqnarray}
v_{\parallel } &\simeq &v_{\parallel }(0)+\frac{\frac{\sum_{k}B_{k}}{B_{0}}%
v_{A}\Omega _{k}\cos \left[ \varphi _{k}-kz(0)\right] \cos \left[ \varphi
_{k}-\Omega _{0}t-kz(0)\right] }{-2\Omega _{0}}  \nonumber \\
&&-\frac{\frac{\sum_{k}B_{k}}{B_{0}}v_{A}\Omega _{k}\cos \left[ \varphi
_{k}-kz(0)\right] \cos \left[ \varphi _{k}+\Omega _{0}t-kz(0)\right] }{%
2\Omega _{0}} \nonumber \\
&&+\frac{\frac{\sum_{k}B_{k}}{B_{0}}v_{A}\Omega _{k}\cos ^{2}\left[ \varphi
_{k}-kz(0)\right] }{\Omega _{0}} \\
&=&v_{\parallel }(0)-\frac{v_{A}}{2}\frac{\sum_{k}B_{k}^{2}}{B_{0}^{2}}\cos %
\left[ \varphi _{k}-kz(0)\right] \cos \left[ \varphi _{k}-\Omega _{0}t-kz(0)%
\right]  \nonumber \\
&&-\frac{v_{A}}{2}\frac{\sum_{k}B_{k}^{2}}{B_{0}^{2}}\cos \left[ \varphi
_{k}-kz(0)\right] \cos \left[ \varphi _{k}+\Omega _{0}t-kz(0)\right]  
\nonumber \\
&&+v_{A}\frac{\sum_{k}B_{k}^{2}}{B_{0}^{2}}\cos ^{2}\left[ \varphi _{k}-kz(0)%
\right] ,
\end{eqnarray}

All the solutions derived below are based on a time scale larger than $\pi/(kv_{th})$, indicating the phase randomization (or heating) is saturated. Unless noted otherwise all the notations below are the same as in CPAWs case. The overall average transverse and parallel velocities at position $z$ can be expressed as: 
\begin{eqnarray}
U_{\perp } &=&-iv_{A}\frac{\sum_{k}B_{k}}{B_{0}}\cos \left[ \varphi
_{k}+kv_{A}t-kz\right] \nonumber \\
&&+\frac{i}{\sqrt{\pi }v_{th}}v_{A}\int_{-\infty }^{\infty }\frac{%
\sum_{k}B_{k}}{B_{0}}\cos \left[ \varphi _{k}-kz(0)\right] e^{-i\Omega
_{0}t}e^{-\left[ \frac{v_{\parallel }(0)}{v_{th}}\right] ^{2}}dv_{\parallel
}(0)  \nonumber \\
&\simeq&-iv_{A}\frac{\sum_{k}B_{k}}{B_{0}}\cos \left[ \varphi _{k}+kv_{A}t-kz%
\right] +iv_{A}\frac{\sum_{k}B_{k}}{B_{0}}e^{-i\Omega _{0}t}A_{k}\cos \left[
\varphi _{k}-kz\right]  \nonumber \\
&=&-iv_{A}\frac{\sum_{k}B_{k}}{B_{0}}\cos \left[ \varphi _{k}+kv_{A}t-kz%
\right] , \label{lbperp} \\
U_{\parallel } &=&\frac{1}{\sqrt{\pi }v_{th}}v_{A}  \nonumber \\
&&\int_{-\infty }^{\infty }\left\{ 
\begin{array}{c}
-\frac{1}{2}\frac{\sum_{k}B_{k}^{2}}{B_{0}^{2}}\cos \left[ \varphi _{k}-kz(0)%
\right] \cos \left[ \varphi _{k}-\Omega _{0}t-kz(0)\right]  \\ 
-\frac{1}{2}\frac{\sum_{k}B_{k}^{2}}{B_{0}^{2}}\cos \left[ \varphi _{k}-kz(0)%
\right] \cos \left[ \varphi _{k}+\Omega _{0}t-kz(0)\right]  \\ 
+\frac{\sum_{k}B_{k}^{2}}{B_{0}^{2}}\cos ^{2}\left[ \varphi _{k}-kz(0)\right]
\end{array}%
\right\} e^{-\left[ \frac{v_{\parallel }(0)}{v_{th}}\right]
^{2}}dv_{\parallel }(0)  \nonumber \\
&\simeq &\frac{v_{A}}{2}\frac{\sum_{k}B_{k}^{2}}{B_{0}^{2}},  \label{lbpara}
\end{eqnarray}

Similarly, $U_{\perp }$ here can be decomposed into $U_{x}$ and $U_{y}$ as
follows: 
\begin{eqnarray}
U_{x} &=&0,  \label{lbx} \\
U_{y} &=&-v_{A}\frac{\sum_{k}B_{k}}{B_{0}}\cos \left[ \varphi _{k}+kv_{A}t-kz%
\right] ,  \label{lby}
\end{eqnarray}

It is of particular interest to find that $U_{x}=0$ at any spatial position $z$, indicating that there is no local transverse bulk motion in the $x$ direction due the lack of wave field forces. Thus, no pseudoheating occurs in the $x$ direction as well, which will be demonstrated straight away. After some manipulations, the asymptotic temperature associated with real heating can be written as: 
\begin{eqnarray}
T_{linx}^{R} &=&T_{liny}^{R}=T_{lin\perp}^{R}=\frac{m_{p}}{2k_{B}\sqrt{\pi }v_{th}}\int_{-\infty }^{\infty
}\left\vert v_{\perp}-U_{\perp}\right\vert ^{2}e^{-\left[ \frac{v_{\parallel }(0)}{%
v_{th}}\right] ^{2}}dv_{\parallel }(0)\simeq T_{0}\left( 1+\frac{m_{i}}{%
2m_{p}}\frac{\delta B_{w}^{2}}{\beta _{p}B_{0}^{2}}\right) , \label{lTRx}\\
T_{lin\parallel }^{R} &=&\frac{m_{p}}{k_{B}\sqrt{\pi }v_{th}}\int_{-\infty
}^{\infty }\left\vert v_{\parallel }-U_{\parallel }\right\vert ^{2}e^{-\left[
\frac{v_{\parallel }(0)}{v_{th}}\right] ^{2}}dv_{\parallel }(0)\simeq
T_{0}\left( 1+\frac{3}{8}\frac{m_{i}}{m_{p}}\frac{\delta B_{w}^{4}}{\beta
_{p}B_{0}^{4}}\right) ,  \label{lTRpara}
\end{eqnarray}

It is interesting that the temperature associated with real heating in the $x$ and $y$ directions is identical, indicating the pitch-angle scattering plays an important role in this process\cite{wangPRL,luPOP}. Finally, by adopting the same ensemble average, we achieve the asymptotic perpendicular temperature incorporating both real and pseudo heating for LPAWs as follows: 
\begin{eqnarray}
T_{linx}^{A+R} &\simeq& T_{0}\left( 1+\frac{m_{i}}{%
2m_{p}}\frac{\delta B_{w}^{2}}{\beta _{p}B_{0}^{2}}\right) ,  \label{lTKx}
\\
T_{liny}^{A+R} &\simeq& T_{0}\left( 1+\frac{3}{2\beta
_{p}}\frac{m_{i}}{m_{p}}\frac{\delta B_{w}^{2}}{B_{0}^{2}}\right) ,~
\label{lTKy}
\end{eqnarray}

Consistent with earlier expectation, Eq.(\ref{lTRx}) and Eq.(\ref{lTKx}) are identical. It is noteworthy that the non-thermal speed in the $x$ direction is zero, whereas it has the same value, $\xi =v_{A}\delta B_{w}/B_{0}$, as the CPAWs case in the $y$ direction. From the expressions above, one could summarize some general characteristics of the non-resonant wave particle scattering: 1) The temperature anisotropy in the perpendicular and parallel directions always exists. The perpendicular heating dominates over parallel heating. 2) No pseudoheating can be observed in the perpendicular direction, where no wave magnetic perturbation presents. In other words, the corresponding overall average transverse velocity equals to zero (see Eq.(\ref{lbx})) in the absence of magnetic field perturbations. If one studies minor ions (e.g., $\alpha$ particles) in the spectrum of finite amplitude low-frequency Alfv\'{e}n waves, it is possible to distinguish real heating from pseudoheating by current available observational data. As we all know, the typical ion cyclotron period of $\alpha$ particles at 1 $AU$ is about 100 $s$ and Alfv\'{e}n speed is about 50 $km/s$, which indicates that measurements with a spatial resolution of about 5000 $km$ and with a temporal resolution of about 100 $s$ is enough for testing this theory for . Both \emph{Wind} and \emph{Helios} spacecrafts provide ion plasma measurements with cadences 90 $s$ and 41 $s$, respectively. 

\section{TEST PARTICLE CALCULATIONS AND DISCUSSIONS}
In this paper, we are interested in minor ions. We adopt $He^{2+}$ as test particle because of its low abundance in the solar wind. The simulation results using test-particle calculations build upon previous works\cite{wangPRL,wangPOP,dongPOP,dongPOP2013}. We discretize the Alfv\'{e}n wave number by $k_{j}=k_{min}+(j-1)\frac{k_{max}-k_{min}}{J-1}$, for $j=1,...,J$, where $k_{min}=k_{1}=1\times 10^{-2}\Omega _{0}/v_{A}$ and $k_{max}=k_{J}=5\times 10^{-2}\Omega _{0}/v_{A}$ ($\Omega _{0}$ here denotes $He^{2+}$ ion gyrofrequency). This range of wave numbers corresponds to $1\times 10^{-2}\Omega _{0}<\omega <5\times 10^{-2}\Omega _{0}$, so that the wave frequencies are much lower than the ion gyrofrequency. The amplitudes of different wave modes are constant and equal to each other. Here we set the value of $\delta B_{w}^{2}/B_{0}^{2}$=0.04. The total number of test particles is $10^{5}$, which are randomly distributed during the time interval $0<\Omega _{0}t<2\pi $ and in the spatial area $0<z\Omega_{0}/v_{A}<3\times 10^{3}$. The initial VFD of $\alpha$ particles is assumed be Maxwellian with a thermal speed at $v_{th}^{\alpha}$=0.07$v_{A}$, thus the cyclotron resonance condition is not satisfied.

\begin{figure}[tbp]
\centering
\includegraphics[scale=0.25]{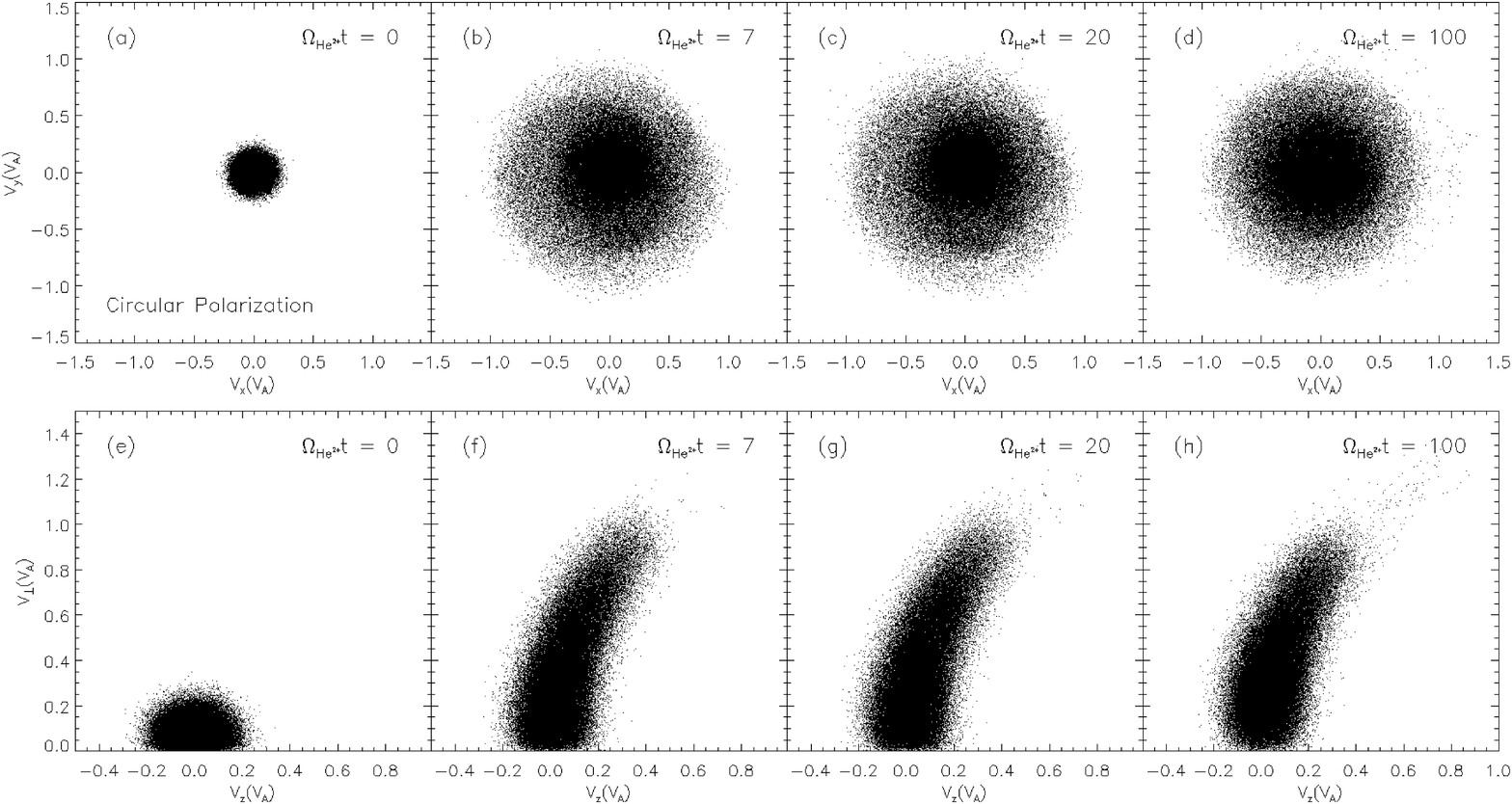}
\caption{Velocity scatter plots of the test particles in the $v_x$-$v_y$
space (first row), and $v_z$-$v_{\perp}$ space (second row) at $\Omega_{0} t$
= 0, 7, 20, 100 for CPAWs.}
\label{cirvel}
\end{figure}

\begin{figure}[tbp]
\centering
\includegraphics[scale=0.206]{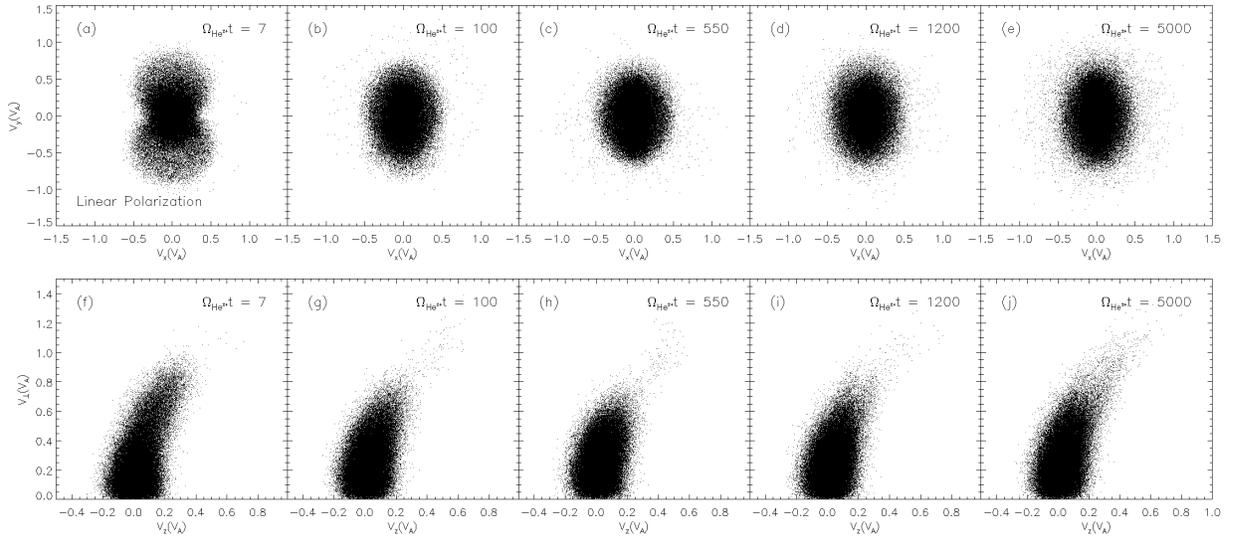}
\caption{Velocity scatter plots of the test particles in the $v_x$-$v_y$
space (first row), and $v_z$-$v_{\perp}$ space (second row) at $\Omega_{0} t$
= 7, 100, 550, 1200, 5000 for LPAWs.}
\label{linvel}
\end{figure}

Fig. \ref{cirvel} shows the velocity scatter plots of the $He^{2+}$ test particles in the $v_x$-$v_y$ space (first row), and $v_z$-$v_{\perp}$ space (second row, $v_{\perp}=\sqrt{v_x^2+v_y^2} \ge 0$) at $\Omega_{0} t$ = 0, 7, 20, 100 in the presence of CPAWs. The simulation results are consistent with the results shown in previous studies\cite{wangPRL,dongPOP}. It is clear that no more heating can be observed after $\Omega_{0} t \approx$ 7, indicating that the heating finishes within one $\alpha$ particle gyroperiod. Fig. \ref{linvel} shows the velocity scatter plots in the presence of LPAWs at $\Omega_{0} t$ = 7, 100, 550, 1200, 5000. Since the initial particle distribution in the velocity phase space is identical, the velocity scatter plots at $\Omega_{0} t$ = 0 in linearly polarized case are not shown. The main difference between Fig. \ref{cirvel} and Fig. \ref{linvel} is the different shapes of the velocity distribution in the $v_x$-$v_y$ space due to their different polarizations [see Eqs.(\ref{lnr}) \& (\ref{cir})]. Inspection of Fig. \ref{linvel} reveals that the ion distribution in the $v_x$-$v_y$ space depends on the direction of linear polarization of the Alfv\'{e}n wave\cite{dongPOP2013}. Although the velocity scatter plots in Fig.\ref{linvel} show almost no difference after $\Omega_{0} t$ = 20, the particle distribution in the phase space still keeps evolving until the system is uniformly heated in the perpendicular direction (see Fig.\ref{linph}). The corresponding time scale is on the order of $\pi/(kv_{th}^{\alpha})$ as indicated in Sec. II.

\begin{figure}[tbp]
\centering
\includegraphics[scale=0.24]{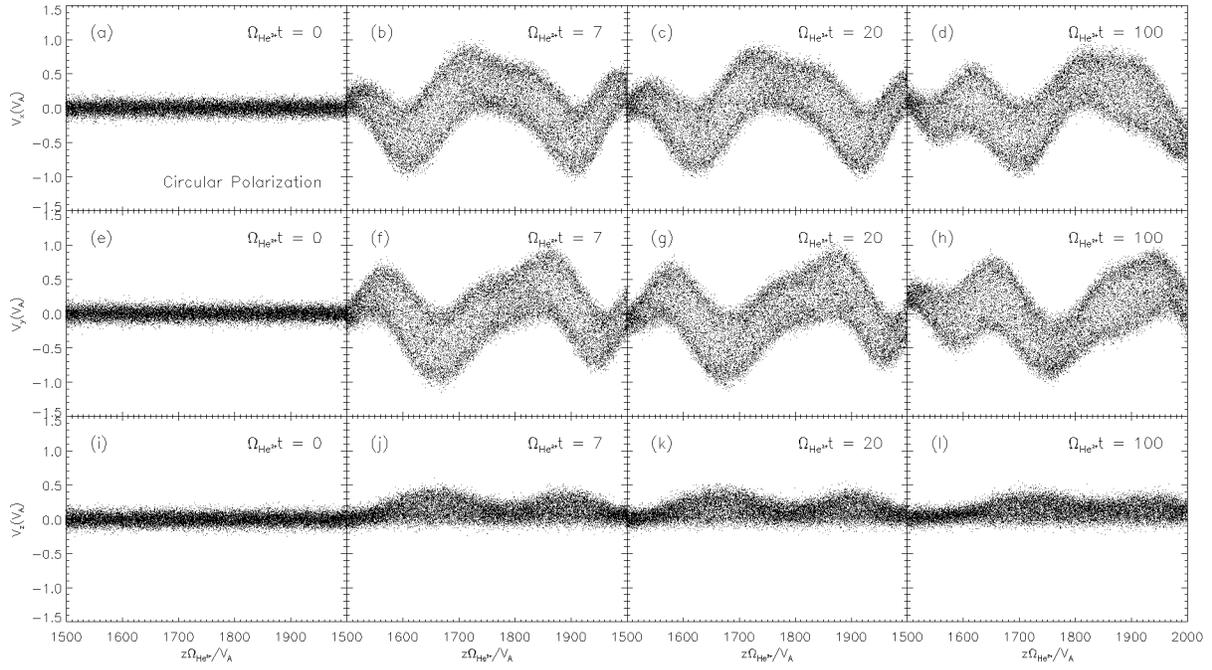}
\caption{Scatter plots of the test particles between 1500 and 2000 $v_A
\Omega_{0}^{-1}$ at different times, $\Omega_{0} t$ = 0, 7, 20, 100 in the $%
v_x-z$ (first row), $v_y-z$ (second row) and $v_z-z$ (third row) for
CPAWs.}
\label{cirph}
\end{figure}

\begin{figure}[tbp]
\centering
\includegraphics[scale=0.235]{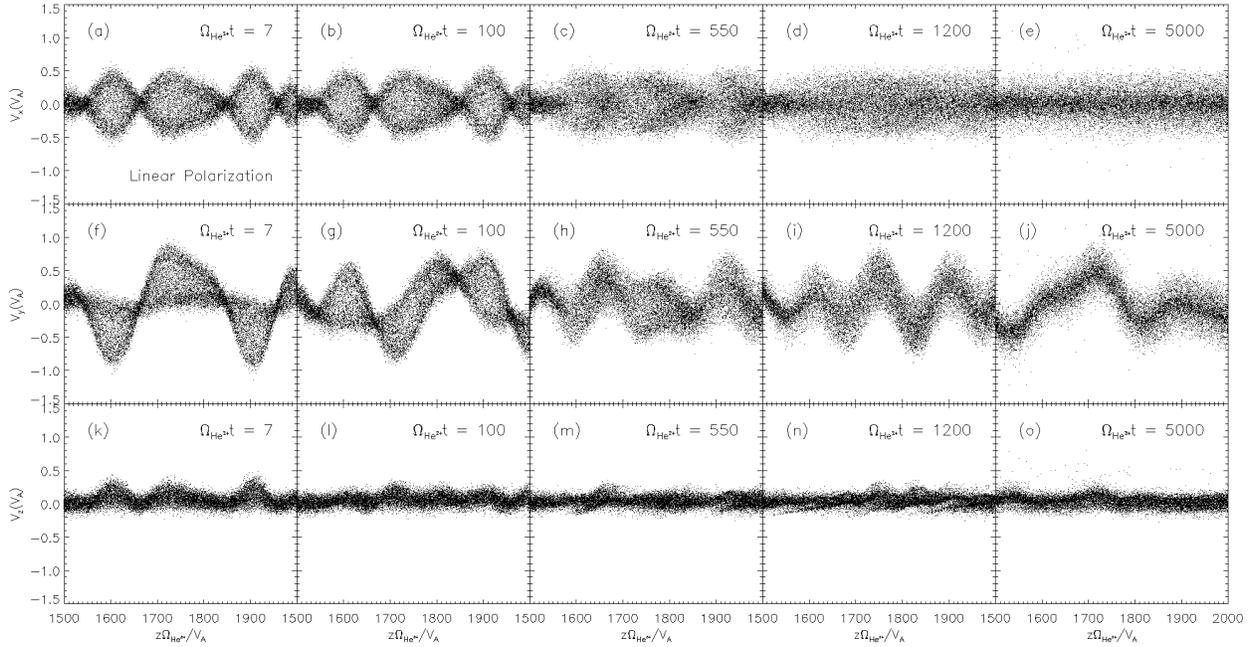}
\caption{Scatter plots of the test particles between 1500 and 2000 $v_A
\Omega_{0}^{-1}$ at different times, $\Omega_{0} t$ = 7, 100, 550, 1200,
5000 in the $v_x-z$ (first row), $v_y-z$ (second row) and $v_z-z$ (third
row) for LPAWs.}
\label{linph}
\end{figure}

The scatter plots of the test particles between 1500 and 2000 $v_A \Omega_{0}^{-1}$ in the phase space in the presence of CPAWs and LPAWs are shown in Fig. \ref{cirph} and Fig. \ref{linph}, respectively. Inspection of Fig. \ref{cirph} reveals the heating finishes within one ion gyroperiod in a spectrum of CPAWs. Although the heating also finishes in one ion gyroperiod in linearly polarized case as indicated in Fig. \ref{linph}; it takes more time for the system to become uniformly heated in the perpendicular direction via phase mixing. The phase space inhomogeneity in the early stage is caused by the nonuniform asymmetric spatial distribution of wave fields. Particles eventually become uniformly heated with the help of non-resonant pitch angle scattering\cite{wuPRL,wuPOP1}. Comparing the first two rows in Fig. \ref{linph}, it is of particular interest to find that only genuine heating shows up in the $v_x-z$ phase space while there exist both genuine and pseudo heating in the $v_y-z$ phase space. This is consistent with our analytic prediction (refer to Eq.(\ref{lTRx}) and Eq.(\ref{lTKx})). The perpendicular pseudoheating can be naturally filtered out in the absence of wave magnetic fluctuations. This phenomenon, however, can never be observed in the presence of CPAWs, in good agreement with the analytic solutions shown in Sec. II. Therefore, the phase space diagram in the linearly polarized case reveals the basic nature of non-resonant heating: non-resonant heating is caused by phase mixing or non-resonant pitch angle scattering, where both genuine and pseudo heating processes take place.

\begin{figure}[tbp]
\centering
\includegraphics[scale=0.4]{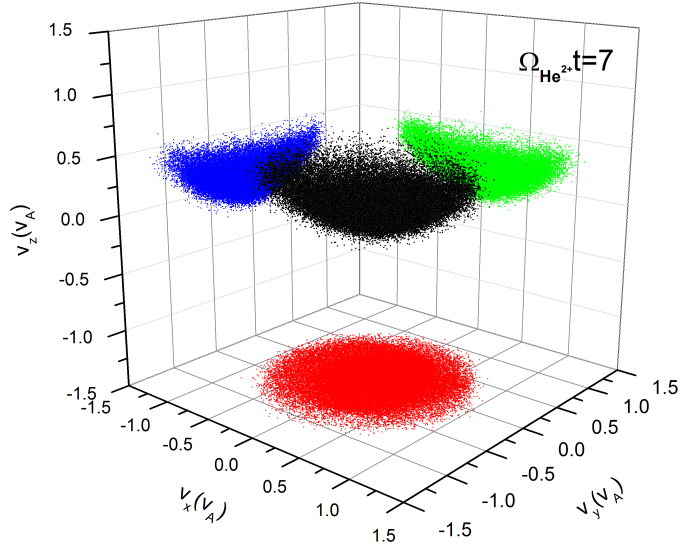}
\caption{(Color online) Velocity scatter plot of the test particles in the
3-D velocity space at $\Omega_{0} t$= 7 for CPAWs.}
\label{3dvel}
\end{figure}

In Fig. \ref{3dvel}, we present a 3-D velocity scatter plot and the corresponding projections. For the case of CPAWs, the ion velocity distributions are gyrotropic in the $v_x-v_y$ space, consistent with the solar wind observation (done by Helios), indicating that ion VDFs are gyrotropic in most cases\cite{Marsch}. However, in LPAWs case, the VDFs are no longer gyrotropic because: firstly, we are interested in the physical picture of phase mixing during the non-resonant wave-particle interaction, so we choose all the wave magnetic fields polarized in the $y$ direction. In reality, it is unrealistic that all the waves in a spectrum are linearly polarized in the same direction. Secondly, in a real space environment, there also exist both obliquely propagating shear Alfv\'{e}n waves and kinetic Alfv\'{e}n waves. In these cases, ion orbit becomes chaotic as long as the threshold condition is satisfied and stochastic heating will take place\cite{chenPOP,WangGRL,Ben,luPOP,Ben1}. When the stochastic heating occurs, the heating efficiency is higher than that of non-resonant scattering. The non-gyrotropic effecters resulted from pesudoheating therefore can be ignored\cite{WangGRL}. We, however, only deal with the parallel propagating shear Alfv\'{e}n waves with finite amplitude. Hence, no stochastic heating is observed. The non-resonant heating process usually occurs much earlier than the stochastic heating (see Fig.2 in Ref.\cite{WangGRL}). Therefore, in this paper we are interested in short-time scale physics. How to accurately quantify the threshold condition for the stochastic heating is still an open question. The non-resonant heating may be correlated to the stochastic heating in some circumstances, which could be an interesting research topic in the future. The non-gyrotropic distribution is of particular importance in some specific space plasma environments, such as the thin boundary layers or reconnection structure. Clear signature of non-gyrotropic energetic electron distributions was found by ISEE 1 and ISEE 2 spacecrafts in the upstream of the Earth's bow shock and has been detected by \emph{in situ} observations of the WIND plasma experiment\cite{nong}. Besides, the appearance of non-gyrotropic ion velocity distributions is well established in the magnetotail, providing another piece of evidence to support the existence of magnetic reconnection
processes\cite{nong}. 

Since the VDFs in a spectrum of CPAWs have been discussed in detail in the previous references\cite{wangPRL,dongPOP,XingLiApJL}, here we only focus on the linearly polarized case. Fig. \ref{dist} shows the normalized VDFs plotted against $v_x$, $v_y$ and $v_z$, at $\Omega_{0} t$ = 0 and 5000 in a spectrum of LPAWs. The solid and plus centered circles denote the normalized VDFs based on the statistics of ions in all spatial range (0 $<z\Omega_0/v_A<$ 3000) at  $\Omega_{0} t$ = 0 and 5000, respectively. The dot centered squares represent the local normalized VDFs around $z\Omega_0/v_A$ = 1800, which is based on the statistics of ions in the spatial range 1790 $<z\Omega_0/v_A<$ 1810. If the spatial range is too narrow, there are insufficient particles to be counted for statistics. In order to compare the width of the VDFs, the associated averaged bulk motion based on the statistics of all ions has been removed in Fig.\ref{dist}. In Fig.\ref{dist} (a), it is clear that the VDFs at $\Omega_{0} t$ = 5000 are broader than that at $\Omega_{0} t$ = 0, indicating the $\alpha$ particles are heated. As indicated in Fig. \ref{linph}, genuine heating is observed only in the $v_x$ direction, therefore the two VDFs are almost the same at $\Omega_{0} t$ = 5000 no matter what statistical spatial ranges are adopted. The slight difference between these two VDFs results from the statistical uncertainties due to the insufficient number of particles in the selected spatial range. The results shown in Fig. \ref{dist} (b) are also consistent with those presented in Fig. \ref{linph}; the VDFs based on the statistics of all the particles are broader than the VDFs in the particles' mean-velocity frame (dot centered square) due to the averaging over wave effects\cite{dongPOP2013,DV}. It is well known that the full width at half maximum (FWHM) of VDFs is proportional to the corresponding temperature. Therefore, the FWHM of the VDF represented by the dot centered square is proportional to the temperature associated with real heating $T_{lnr}^{R}$, while the FWHM of the VDF denoted by the plus centered circle is proportional to $T_{lnr}^{A+R}$. Inspection of Fig .\ref{dist} (c) reveals that it is slightly heated in the parallel direction, which also agrees well with the analytic solution shown in Sec. II.

\begin{figure}[tbp]
\centering
\includegraphics[scale=0.33]{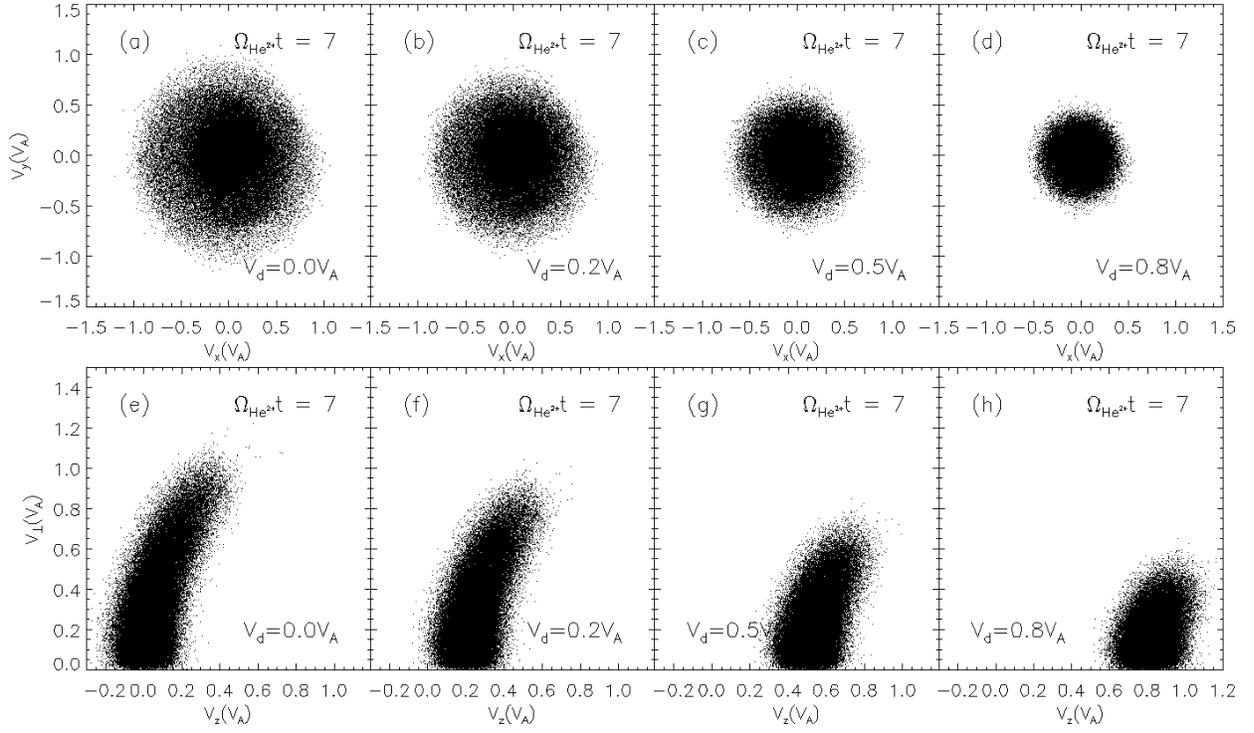}
\caption{Velocity scatter plots of the test particles in the $v_x$-$v_y$ space (first row), and $v_z$-$v_{\perp}$ space (second row) for normalized drift velocity $v_d/v_A$ = 0, 0.2, 0.5, 0.8 at $\Omega_{0} t$ = 7 in circularly polarized case.}
\label{cirveld}
\end{figure}

\begin{figure}[tbp]
\centering
\includegraphics[scale=0.4]{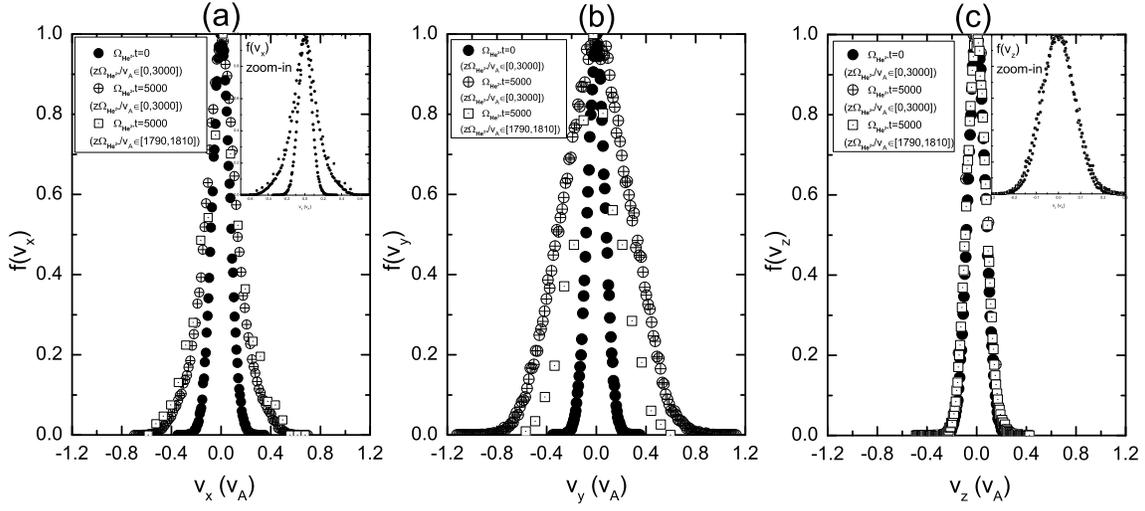}
\caption{The normalized VDFs plotted against (a) $v_x$, (b) $v_y$ and (c) $v_z$, at $\Omega_{0} t$ = 0 and 5000 in a spectrum of LPAWs. The solid and plus centered circles denote the normalized VDFs based on the statistics of ions in all spatial range at $\Omega_{0} t$ = 0 and 5000, respectively. The dot
centered squares represent the local normalized VDFs based on the statistics of ions in the spatial range 1790 $<z\Omega_0/v_A<$ 1810. In order to compare the width of the VDFs, the averaged bulk velocity based on the statistics of ions in the corresponding spatial range has been removed from the plots.}
\label{dist}
\end{figure}

Since the $\alpha$ particle drift (or differential flow) speed plays an important role in the ion perpendicular heating\cite{JK2,Ben1,SB1,SB2,Marschhe}, we investigate the effect of normalized drift velocity, $v_d/v_A$, on the $\alpha$ particle perpendicular heating in process of non-resonant wave particle scattering. Fig. \ref{cirveld} shows velocity scatter plots for several different normalized drift velocity $v_d/v_A$= 0.0, 0.2, 0.5, 0.8 in a spectrum of CPAWs. The simulation results indicate the larger the drift speed is, the smaller the perpendicular heating is. This tendency is consistent with the $Wind$ observations\cite{JK2}. In real fast solar wind conditions, the $\alpha$ particles drift relative to the solar wind protons, which indicates the Alfv\'{e}n wave electric field seen by alpha particles decrease in the case of parallel propagation\cite{Ben}. 

If one wants to partially reproduce the main features of the observed solar wind proton VFDs, kinetic Alfv\'{e}n waves (KAWs) and ion acoustic waves (IAWs) play key roles. $Li~et~al.$ employed a 1-D test particle code together with a linear Vlasov numerical solver to investigate the effect of a nearly perpendicular propagating kinetic Alfv\'{e}n wave on the shape of
the proton VDFs. They found that the KAWs are able to generate a secondary proton beam in collisionless plasmas such as the solar wind by Landau resonance. The proton beam occurs in the direction along the background magnetic field $B_0$\cite{XingLiApJL}. $Araneda~et~al.$ used Vlasov theory and a 1-D hybrid code to show that the ion acoustic waves (IAWs) and transverse waves driven by Alfv\'{e}n-cyclotron wave parametric instabilities can lead to preferential heating and acceleration of heavy ions such as $\alpha$ particles, and formation of a field aligned isolated proton beam\cite{jaaprl2008,jaaprl2009}. Rather remarkably, our analytic solution could also capture the feature of the preferential heating of heavy minor ions in the solar wind to a certain degree. In our case, however, it is almost impossible to get the double-beam structure by parallel propagating shear Alfv\'{e}n wave via test particle simulations. On the other hand, for the heavy minor ions such as $He^{2+}$, because of their larger mass, ions do not become significantly trapped and therefore there is no obvious double-beam structure \cite{jaaprl2009}. Further investigation of non-resonant wave-proton interaction requires self-consistent approaches such as hybrid and PIC simulations.

Finally, in order to show that the non-resonant heating includes both the (irreversible) genuine and (reversible) pseudo heatings, we consider that the amplitude of each wave mode changes gradually with time such that $\delta B_{w}^{2}(t)=\sum_{k}B_{k}^{2}(t)=\epsilon (t)B_{0}^{2}$ in circularly polarized case, where
\begin{center}
$\epsilon (t)=\left\{ 
\begin{array}{ccc}
\epsilon _{0}, & if & t\leq t_{1}, \\ 
\epsilon _{0}e^{-(t-t_{1})^{2}/\tau ^{2}}, & if & t>t_{1}.%
\end{array}%
\right. $
\end{center}

\begin{figure}[tbp]
\centering
\includegraphics[scale=0.6]{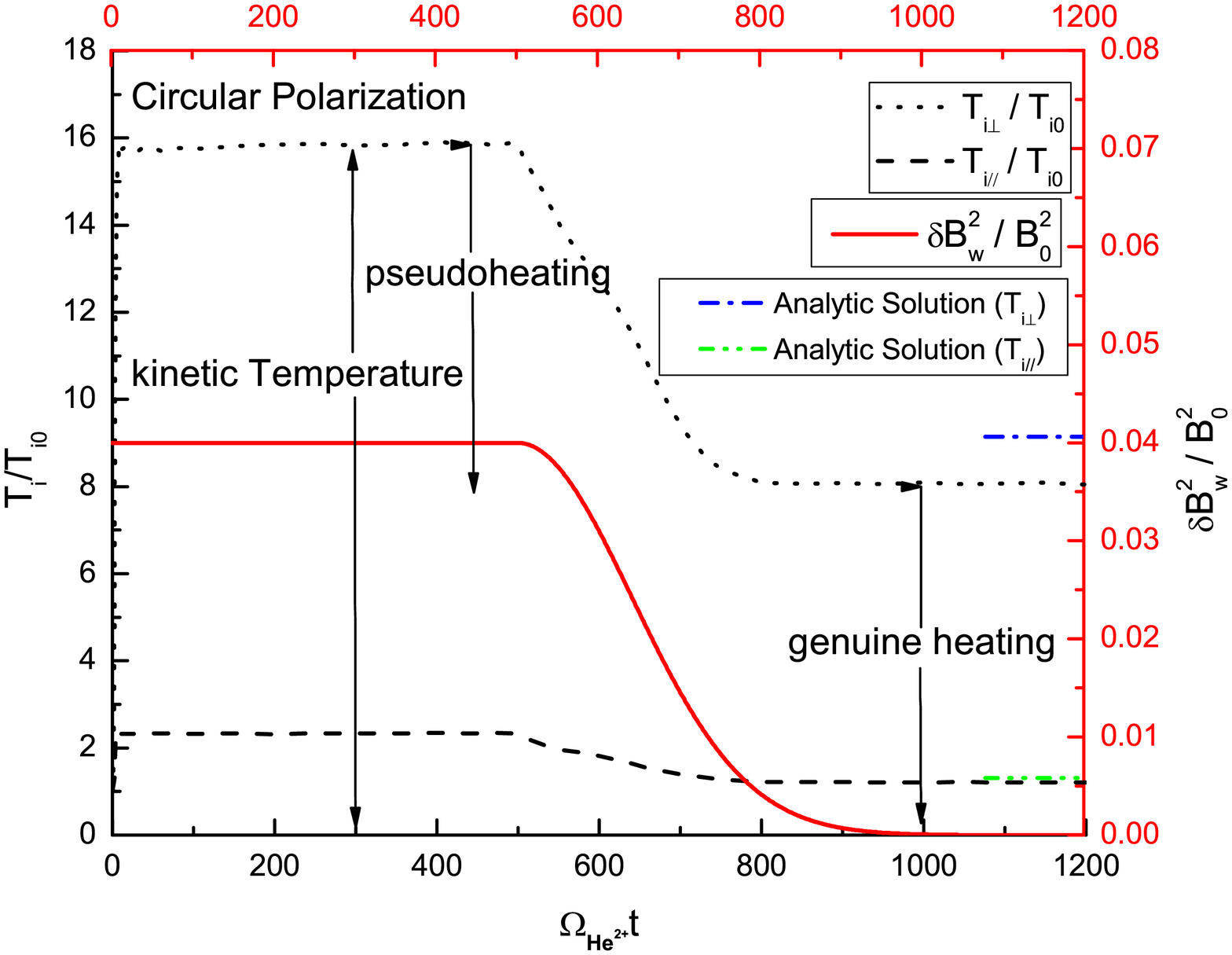}
\caption{(Color online) Non-resonant heating in temporally evolved wave fields for CPAWs.}
\label{tempdiff}
\end{figure}

Similar to the previous studies\cite{wangPOP,dongPOP2013}, we set $t_{1}$=500$\Omega _{0}^{-1}$, $\tau $=200$\Omega _{0}^{-1}$, and $\epsilon _{0}$=0.04. The temporal evolution of the wave fields and the temperature of $\alpha$ particles are shown in Fig. \ref{tempdiff}. Initially the newborn $\alpha$ particles are picked up by the turbulent Alfv\'{e}n waves. They are heated via the non-resonant pitch-angle scattering or phase mixing. At this stage, the temperature is associated with both genuine and pseudo heating. After the wave diminishing, the ions do not return to the unheated state but are heated with respect to their initial temperature. Hence, this process is partially irreversible and the remaining temperature is associated with genuine heating.

The non-resonant wave particle interaction may play an important role in various space environments, especially in the solar corona. Recently, Huang \emph{et al.} discovered the so-called down loop structure, where the temperature decreases with height, in the solar corona (especially in the low latitude closed field regions) by using the Differential Emission Measure Tomography (DEMT) technique\cite{Huang}. Some future work may involve studying the relation between the non-resonant heating and the down loop structure by varying the plasma parameters, such as plasma $\beta$ and Alfv\'{e}n wave amplitude, provided by the newly developed and validated global MHD wave and coronal heating model\cite{Igor,Bart}. In other words, it requires the coupling between the test particle model and the Alfv\'{e}n turbulence MHD model. One possible explanation for the down loop structure is that for CPAWs, the non-resonant heating finishes in a short time scale that is on the order of one ion cyclotron period, thus the Alfv\'{e}n waves damp before they have a chance to propagate upward in the low latitude closed field regions.

\section{CONCLUSIONS}

In summary, the analytic solutions of non-resonant heating, especially for a spectrum of LPAWs, are presented. The heating is dominant in the direction perpendicular to the background magnetic field. The perpendicular heating, $\eta=(T_{i\perp}^R-T_{i0\perp}^R)/T_{i0\perp}^R$ (where $T_{i0\perp}^R$ and $T_{i\perp}^R$ are the perpendicular temperature of species $i$ before and after genuine heating, respectively), in the spectrum of CPAWs is about a factor of two stronger than that of LPAWs. The analytic solutions reproduce the preferential perpendicular heating of solar wind minor heavy ions to a certain degree, in good agreement with the test $\alpha$-particle simulation results. We also show that the drift (or differential flow) speed, $v_d$, has an effect on reducing the ion perpendicular heating, which is consistent with observations. The pseudoheating is associated with wave fluctuations that contribute to the non-thermal broadening of VDFs due to the wave field forces. Therefore, no pseudoheating heating is observed in the perpendicular direction where no wave magnetic fluctuations present. It takes much longer for a system to be uniformly heated in the perpendicular direction in a spectrum of LPAWs. The evolution of the particle distribution in the phase space reveals that the non-resonant heating is caused by phase mixing or non-resonant pitch angle scattering.

From the phase space diagram, it is relatively easy to distinguish real heating from pseudoheating by our model simulations. It is also possible to test this theory for solar wind minor ions, such as $\alpha$ particles, at 1 $AU$ based on the current available data from \emph{Wind} and \emph{Helios} spactcrafts. For the near solar region, however, \emph{in situ} observations from high spatial and temporal resolution space instruments is required. The upcoming NASA Solar Probe Plus mission, which will approach to within 9.5 solar radii of the center of the Sun and sample the sub-Alfv\'{e}nic corona directly. This mission may provide a great opportunity to test this theory in the sub-Alfv\'{e}nic corona. Hopefully this paper could draw attention to the space and solar observation community.

\vskip 10mm

\noindent {\large \textbf{Acknowledgments:}} C.F. Dong appreciates many fruitful discussions with Prof. C.B. Wang and Prof. C.S. Wu at USTC; Prof. L.A. Fisk, Prof. Y.Y. Lau, Prof. J.C. Kasper and Dr. I.V. Sokolov at University of Michigan; Dr. L. Dai. at University of Minnesota; Dr. D. Winske, Dr. S.P. Gary, Dr. M.M. Cowee, and Dr. X.R. Fu at LANL; Prof. P.M. Bellan at Caltech and Dr. D. Verscharen at the University of New Hampshire. C.F. Dong also would like to thank the anonymous referees' helpful comments and suggestions.

\end{document}